\documentclass[twocolumn,superscriptaddress]{revtex4}

\usepackage{graphicx} 

\begin{document}

\title{Superlattice Structures of Graphene based Nanoribbons}

\author{H. Sevin\c{c}li}
\affiliation{Department of Physics, Bilkent University, Ankara 06800, Turkey}
\author{M. Topsakal}
\affiliation{UNAM-Institute of Materials Science and Nanotechnology, Bilkent University, Ankara 06800, Turkey}
\author{S. Ciraci} 
\email{ciraci@fen.bilkent.edu.tr}
\affiliation{Department of Physics, Bilkent University, Ankara 06800, Turkey}
\affiliation{UNAM-Institute of Materials Science and Nanotechnology, Bilkent University, Ankara 06800, Turkey}

\begin{abstract}
Based on first-principles calculations we predict that periodically repeated junctions of armchair graphene nanoribbons of different widths form superlattice structures.
In these superlattice heterostructures the width and the energy gap are modulated in real space and specific states are confined in certain segments.
Orientation of constituent nanoribbons, their width and length, the symmetry of the junction are the structural parameters to engineer electronic properties of these quantum structures. 
Not only the size modulation, but also composition modulation, such as periodically repeated, commensurate heterojunctions of BN and graphene honeycomb nanoribbons result in a multiple quantum well structure. 
We showed that these graphene based quantum structures can introduce novel concepts to design nanodevices.
\end{abstract}

\pacs{73.22.-f, 72.80.Rj, 75.70.Ak}

\maketitle

Recent advances in materials growth and control techniques have
made the synthesis of the isolated, two-dimensional (2D) honeycomb
crystal of graphene possible \cite{novo,zhang,berger}. Owing to its
unusual electronic energy band structure leading the charge
carriers resembling massless Dirac Fermions, graphene introduced
novel concepts and initiated active research \cite{kats,geim1,geim2,ozyilmaz,cohen_hm}.
For example, quasi 1D graphene nanoribbons exhibit interesting size
and geometry dependent electronic and magnetic
properties \cite{ribbon1,ribbon2,ribbon3,cohen_gap}.

In this letter, based on extensive first-principles, as well as empirical tight-binding (ETB) calculations we showed that periodically repeating heterojunctions made of graphene nanoribbons of different widths can form stable superlattice structures. 
These superlattices are unique, since both size and the energy band gap are modulated in real space. 
As a consequence, in addition to the propagating states, specific states are confined in certain regions. 
Confinement increases and turns to a complete Mott's localization when the extent of the nanoribbons with different widths increases. 
Widths, lengths, chirality of constituent nanoribbons and the symmetry of the junction provide variables to engineer quantum structures with novel electronic and transport properties. 
We also showed that multiple quantum well structures can be formed through composition modulation of commensurate heterojunction of BN honeycomb and graphene nanoribbons both having the same width.
In particular, our transport calculations demonstrate that specific finite size quantum structures can operate as resonant tunnelling double barrier (RTDB) with resonances, characteristics of quantum well states.
Since nanoribbons having width less than $\sim$10 	nm with uniform and patterned edges have been produced \cite{dai}, present results related with graphene based nanoribbons and their heterojunctions of diverse geometry can, in fact, be
a candidate for a class of nanodevices with a richness of novel properties. 

We have performed first-principles plane wave calculations \cite{vasp1,vasp2} within density functional theory (DFT) \cite{dft1,dft2} using projector augmented wave (PAW) potentials \cite{paw}.
The exchange correlation potential has been approximated by Generalized Gradient Approximation (GGA) using PW91 functional\cite{pw91} both for spin-polarized and spin-unpolarized cases. 
All structures have been treated within supercell geometry using the periodic boundary conditions. 
A plane-wave basis set with kinetic energy cutoff of 500 eV has been used. In the self-consistent potential and total energy calculations the Brillouin zone (BZ) is sampled by (1$\times$1$\times$35) special \textbf{k}-points for nanoribbons. 
This sampling is scaled according to the size of superlattices. 
All atomic positions and lattice constants are optimized by using the conjugate gradient
method where total energy and atomic forces are minimized. 
The convergence for energy is chosen as 10$^{-5}$ eV between two steps, and the maximum force allowed on each atom is 0.02 eV/\AA.

\begin{figure}
\begin{center}
\includegraphics[scale=0.42]{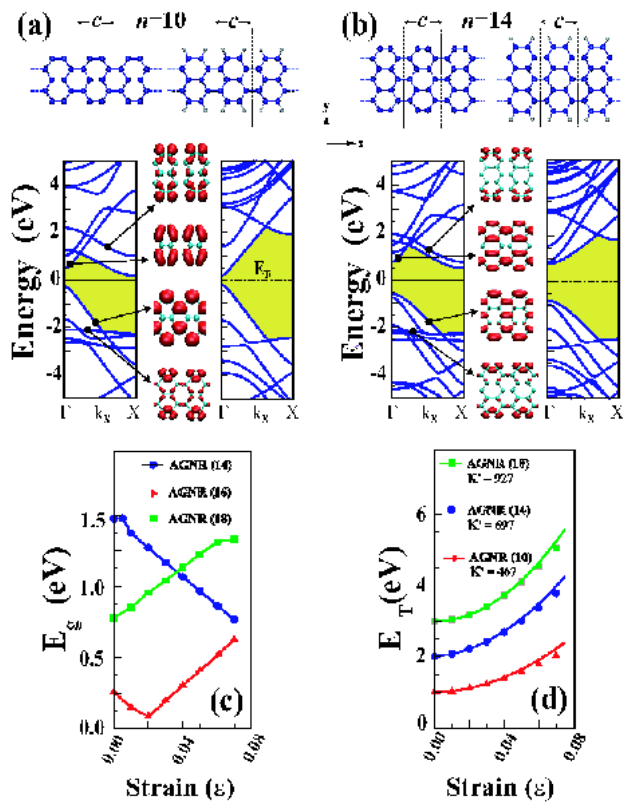}
\caption{(a) Bare and hydrogen terminated AGNR(10): Atomic geometry, electronic band structure and isosurface charge densities of edge and ``uniform" states. The primitive unit cell is delineated with dashed lines and includes $n$=10 carbon atoms. 
Carbon and hydrogen atoms are shown by large and small balls. 
(b) Same for AGNR(14).
(c) Variation of band gaps $E_{g}$ of AGNR, with the tensile strain, $\epsilon$. 
(d) Variation of the total energy of AGNR with $\epsilon$ and its second derivative with respect to $\epsilon$, i.e. $\kappa^{'}$. 
All data in this figure are calculated by first-principles method.}
\label{fig:1}
\end{center}
\end{figure}

Here we consider the bare and hydrogen-terminated armchair graphene nanoribbons, AGNR($n$), $n$ being the number of carbon atoms in the primitive unit cell.
All AGNR($n$) are semiconductors and their band gap, $E_g$, vary with $n$. 
It is small for $3n/2-1$, but from $3n/2$ to $3n/2+1$ it increases and passing through a maximum it becomes again small at the next minimum corresponding to $3n/2+2$  \cite{cohen_gap}.
As $E_{g}$ oscillates with $n$ its value decreases eventually to zero as $n\rightarrow$ $\infty$.
AGNR($n$) is nonmagnetic.
In Fig.~\ref{fig:1} (a) and (b) the band gap values of bare AGNR(10) and AGNR(14) are 0.44 and 1.10 eV, respectively. 
According to isosurface plots one distinguish the relatively uniform states (having their charge uniformly 
distributed across the nanoribbon) from the edge states (having their charge accumulated at both edges of the ribbon). 
Owing to the interaction between two edges of the narrow ribbon, the bands of edge states split.
Upon termination of the carbon dangling bonds by hydrogen atom, the edge states of AGNR(10) and AGNR(14) disappear, and their energy band gaps change to 0.39 eV and 1.57 eV, respectively.
Since hydrogen terminated AGNR's are stable at ambient conditions, in the rest of the paper we consider their superlattices unless stated otherwise.

The rate of change of the band gap with strain, $\epsilon$ in Fig.~\ref{fig:1}(c), i.e. $\partial E_{g}/\partial \epsilon$ is significant and its sign changes with $n$.
The mechanical properties of AGNR($n$) are also important for any device application.
Here we calculated the variation of total energy with strain, $\epsilon$ and its second derivative with respect to strain, $\kappa^{'}= \partial^{2} E_{T}/\partial \epsilon ^{2}$. Results summarized in Fig.~\ref{fig:1} (d) indicate that AGNR's are quasi 1D, stiff materials.
For example, $\kappa^{'}=$697 eV/cell for AGNR(14).  These values can be compared with
$\kappa^{'}$=127 eV/cell calculated for linear carbon chain \cite{tongay} and confirm the robustness of heterostructures.

\begin{figure}
\begin{center}
\includegraphics[scale=0.42]{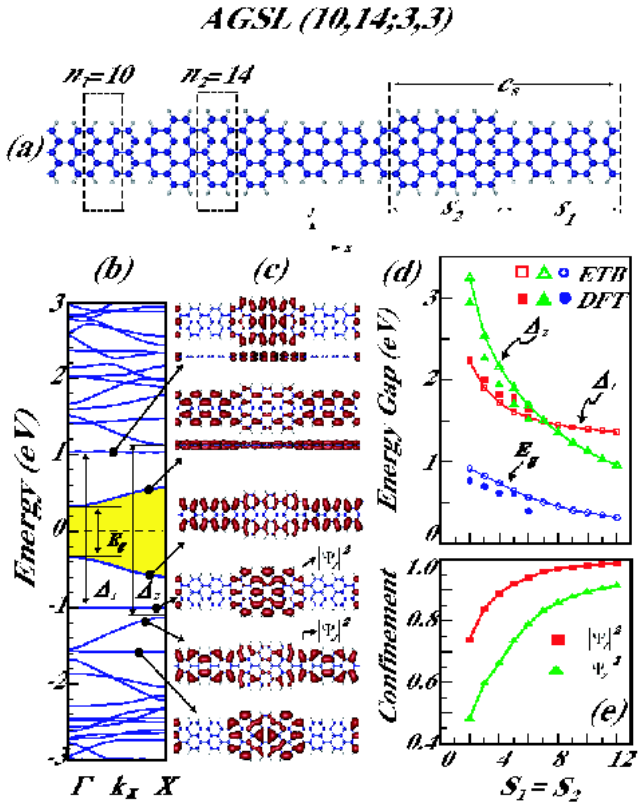}
\caption{(a) Atomic structure of AGSL($n_{1}=10$, $n_{2}=14$; $s_{1}=3$, $s_{2}=3$).
The superlattice unit cell and primitive unit cell of 
each segment are delineated. (b) Band structure with flat bands corresponding to
confined states. (c) Isosurface charge density of propagating and
confined states. (d) Variation of various superlattice gaps with
$s_{1}=s_{2}$. (e) Confinement of states versus $s_{1}=s_{2}$.} 
\label{fig:2}
\end{center}
\end{figure}

The remarkable properties of graphene ribbons discussed above,
especially their band gap rapidly varying with their width,
suggest that the heterostructures formed by the segments of AGNR's with different $n$ may have interesting functions.
Here the crucial issues to be addressed are how the electronic structure will be affected from the discontinuity of crystal potential at the junction, 
what the character of the band discontinuity and the resulting band alignment will be,
and whether these discontinuities will result in confined states. 
To this end we first consider a superlattice AGSL($n_{1}$,$n_{2}$;$s_{1}$,$s_{2}$) made by the segments of AGNR($n_{1}$) and AGNR($n_{2}$). 
Here, $s_{1}$ and $s_{2}$ specify the length of segments in terms of the numbers of their unit cell.
Fig.~\ref{fig:2} shows the superlattice AGSL(10,14;3,3). It is made by periodically deleting one row of carbon atoms at both edges of AGNR(14) to form periodic AGNR(10)/AGNR(14) junction. 
Upon junction formation, dramatic changes occur in the band structure of this superlattice. While highest (lowest) valance (conduction) band is dispersive and their states propagate across the superlattice, the second valence and conduction bands are flat and their states are confined to the wider part of AGSL(10,14;3,3) consisting of a segment of AGNR(14).
These latter flat band states are identified as \emph{confined} states.

The superlattice AGSL(10,14;3,3) in Fig.~\ref{fig:2}(a) can be viewed as if a thin slab with periodically modulated width in the $xy$-plane.
The electronic potential in this slab is lower ($V<$0) than outside vacuum ($V=0$) and thus is modulated in real space.
Normally, electrons in this thin potential slab propagate along the $x-$axis. 
If the junction between wide segment [AGNR(14)] and narrow segment [AGNR(10)] is sharp, some electronic states in the wide segment happen to reflect from sharp discontinuities at the junction and eventually become confined.
The itinerant states are one dimensional, and the band gap and width of this superlattice are modulated in real space.
On the other hand, the atomic arrangement and lattice constants at both sides of the junction are identical; the hetero character concerns only the different widths of the ribbons at different
sides.
According to this simple picture, the larger $\Delta n =n_{2}-n_{1}$ (and also the larger $s_1$ and $s_2$, or the sharper the junction), the stronger becomes the confinement.
As justified in further discussions, these arguments are relevant for superlattices with long constituent segments having lengths larger than de Broglie wavelengths.
Note that AGNR's can be constructed by using two different unit cells (i.e. those consisting of complete hexagons or incomplete hexagons). 
For the wider parts of the superlattice the one with complete hexagons is preferred in order to prevent a lonely carbon atom at the interface. 
For AGSL superlattices with reflection symmetry the narrow region is made by unit cells having complete or incomplete hexagons depending on whether $(n_2-n_1)/4$ is an even or odd number, respectively.

In addition to $n_{1}$, $n_{2}$, $s_{1}$, $s_{2}$, the symmetry of the junction, $\Delta n=n_{2}-n_{1}$, even-odd disparity of $n_{1}/2$ and $n_{2}/2$, the type of the interface
between two different ribbons and the overall shape of the superlattice (namely sharp rectangular, smooth wavy, or sawtooth like) influence the electronic properties. 
As shown in Fig.~\ref{fig:2} (d), as $s_{1}$=$s_{2}$ increases, various superlattice band gaps decrease. 
In terms of the weight of the states in segment $s_2$, i.e. 
$\int_{s_{2}}|\Psi({\bf r})|^{2} d{\bf r}$, 
we see that the confinement of states increase with increasing $s_{1}$=$s_{2}$; 
but it disappears for $s_{1}$=$s_{2}$=1.
Differing of $s_{1}$ from $s_{2}$ causes $E_{g}$ and the band structure to change. 
For example, AGSL(10,14;$s_1$,$s_2$) has $E_{g}$=0.66, 0.48, 0.38, 0.32, 0.27 eV for $s_2$=3, but $s_{1}$=3, 4, 5, 6, 7, respectively. 
Conversely, $E_g$=0.72, 079, 0.83, 0.84 eV for $s_1$=3, but $s_2$=4, 5, 6, 7, respectively. 
On the other hand, the energy of the flat-band states confined to $s_{2}$ and their weight are 
practically independent of $s_{1}$.
Variation of $\Delta n=n_{1}-n_{2}$ results in a wide variety of electronic structures. 
For example, in contrast to AGSL(10,14;3,3) the highest valence and lowest conduction bands of
AGSL(10,18;3,3) are flat bands with $E_{g}$=0.70 eV; 
dispersive bands occur as second valence and conduction band having a gap of 1.19 eV between them. The situation is even more complex for AGSL(10,22;3,3): While the first and the second bands are dispersive in both valence and conduction bands with a direct band gap 0.23 eV, flat bands occur as third bands with a gap of 2.61 eV between them. 
The states associated with the flat bands near the Fermi level are confined at the wider part of the superlattice. 
It should be noted that the band gap of nanoribbons, $E_{g}$ is underestimated by the DFT calculations\cite{cohen_gw}.
Since we consider structures which have already a band gap, its actual value does not affect our discussion in any essential manner, but an enhancement of predicted properties (such as the
extent of confinement) can be expected.

\begin{figure}
\begin{center}
\includegraphics[width=8cm]{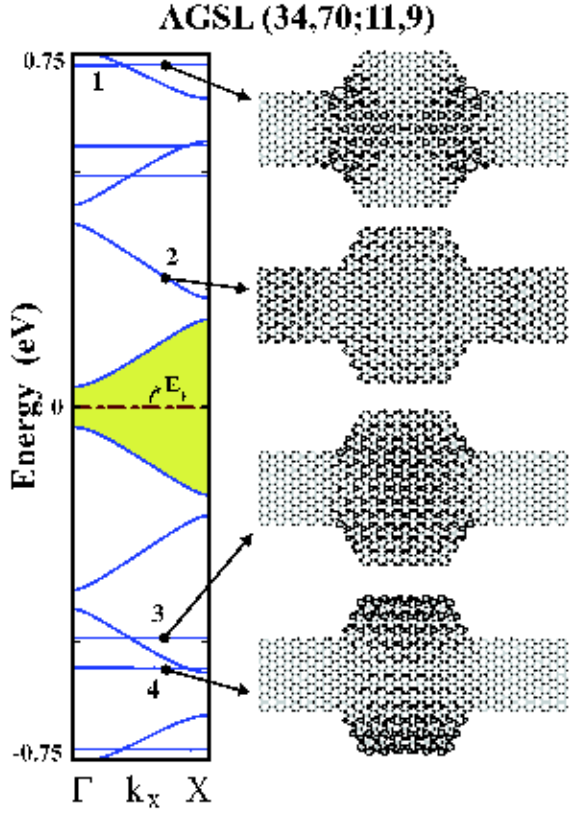}
\caption{Energy band structure of the AGSL(34,70;11,9) superlattice and the charge densities of selected bands.
As seen clearly, states associated with flat bands 1,3, and 4 are confined but the state with dispersive band indicated by 2 is propagating.
Calculations have been performed using ETB method.} 
\label{fig:tb}
\end{center}
\end{figure}

The above trends corresponding to small $n_{1}$ and $n_{2}$ become even more interesting when $n_{1}$ and $n_{2}$ increase.
The electronic band structure of AGSL($n_{1}$,$n_{2}$;$s_{1}$,$s_{2}$) with $n_{1}$=42 
or 82, but $n_{2} > n_{1}$ and $s_{1}$=$s_{2} \geq$ 3 calculated using ETB method shows that for small $\Delta n$, confinement is weak and bands are dispersive, but confinement increases as $\Delta n$ increases. 
Interestingly, $E_{g}$ of AGSL($n_{1}$=42,$n_{2}$;3,3) is, respectively, 0.46, 0.12, 0.49 and 0.04 eV for $n_{2}$=46, 50, 54 and 58. 
In Fig.~\ref{fig:tb} we demonstrate that the confined states can occur not only in narrow (small $n_1$ and $n_2$), but also in wide superlattices having significant modulation of the width.
In ETB method used here, the Bloch states having band index $l$ and wavevector {\textbf{k}} are expressed in terms of the linear combination of the orthonormalized Bloch sums $\chi_i(\mathbf{k},\mathbf{r})$ constructed for each atomic orbital
p$_z$ localized at different carbon atoms, $i$,
with the proper phase of \textbf{k}, namely $\psi_l(\mathbf{k},\mathbf{r})=\sum_i a_{i,l}(\mathbf{k})\chi_i(\mathbf{k},\mathbf{r})$.
Accordingly, the contribution of the orbital at site $i$ to the normalized charge density of $\psi_l(\mathbf{k},\mathbf{r})$ is given by
$\rho_{i}=\sum_l|a_{i,l}(\mathbf{k})|^2$.
In Fig.~\ref{fig:tb}, $\rho_{i}$ is scaled with the radius of circles located at atomic site $i$.

Electronic structure is also strongly dependent on whether the geometry of superlattice is symmetric  (having a reflection symmetry with respect to the superlattice axis along the  $x$-direction), or saddle (one side is straight, other side is periodically caved) all having the same $\Delta n$.
While the saddle structure of AGSL(10,18;3,3) has largest direct gap between dispersive conduction and valence bands, its symmetric structure has smallest gap, but largest number of confined states. 
Horn-like smooth connection between wide and narrow segments may give rise to adiabatic electron transport and focused electron emission \cite{tekman}.

\begin{figure}
\begin{center}
\includegraphics[width=8cm]{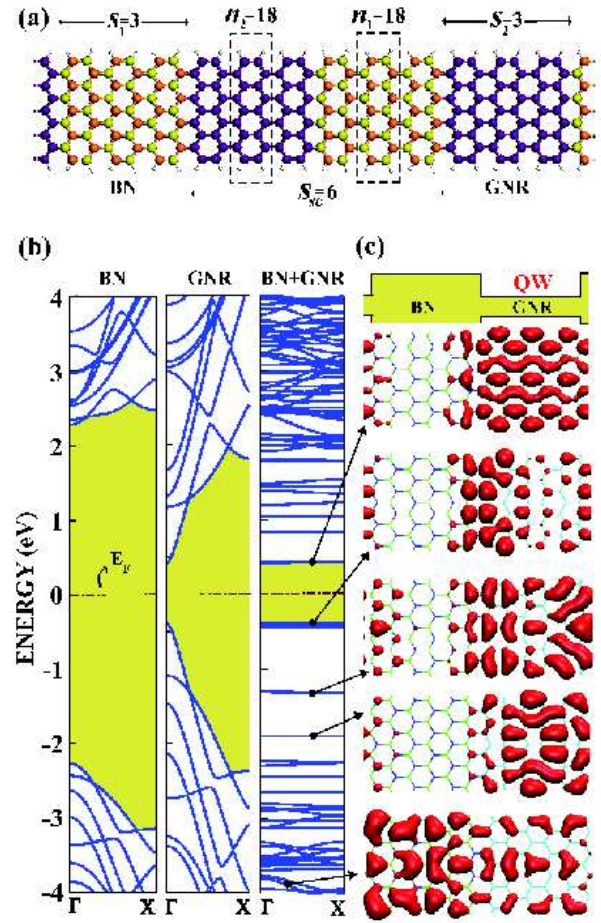}
\caption{One-dimensional superlattice structure formed from the junction of BN and graphene armchair nanoribbons. (a) Atomic structure and superlattice parameters.
(b) Band structures of constituent BN and graphene armchair nanoribbons having 18 atoms in their unit cells and the band structure of the superlattice BN(18)/AGNR(18) each segment having 3 unit cells ($s_1=s_2=3$). (c) Energy band diagram in real space forming multiple quantum wells, QW, in graphene segments (zones). Isosurface charge densities of states confined to QW's and propagating states are presented for selected bands.} 
\label{fig:BN}
\end{center}
\end{figure}

The confined states have been treated earlier in commensurate or pseudomorphic junctions of two different semiconductors, which form a periodically repeating 2D superlattice structure \cite{esaki}. 
These superlattices have grown layer by layer and formed a sharp lattice matched interface. Owing to the band discontinuities at the interface, they behave as if a multiple quantum well structure according to the Effective Mass Theory \cite{bastard}.
Two dimensional conduction band electrons (valence band holes) confined to the wells have displayed a number of electronic and optical properties \cite{ciraci1,ciraci2,ciraci3,ciraci4}.
In Fig.~\ref{fig:BN}, we present a 1D analog of the 2D semiconductor superlattices through compositional modulated nanoribbons. 
BN honeycomb ribbon and graphene ribbon of the same width are lattice matched and can form superlattices with multiple quantum well structure with confined states. 
We considered a periodic junction of the segment of armchair BN ribbon with $n_{1}$=18 and $s_{1}$=3 to the segment of armchair graphene with $n_{2}$=18 and $s_{2}$=3 to form a superlattice structure. 
While periodic BN and graphene ribbons by themselves have band gaps of $\sim$5 eV and 0.8 eV, respectively, the band gap of BN/AGNR(18) is only 0.8 eV indicating a normal (type-I) band alignment.
Under these circumstances, a state propagating in one segment becomes confined if its energy coincides with the band gap of the adjacent BN segment.
In Fig.~\ref{fig:BN}~(b), the dispersive minibands and non-dispersive quantum well states are clearly seen. That these quantum well states are confined in the graphene zone (which has small band gap as compared to the band gap of BN ribbon) are demonstrated by isosurface plots of charge densities. In contrast to the confined states, the propagating states have charge  densities in both graphene and BN zones of the superlattice. 
This is another class of heterostructure obtained from graphene based ribbons and 
their functions can even be advanced by implementing the size modulation in addition to the compositional one.

\begin{figure}
\begin{center}
\includegraphics[scale=0.42]{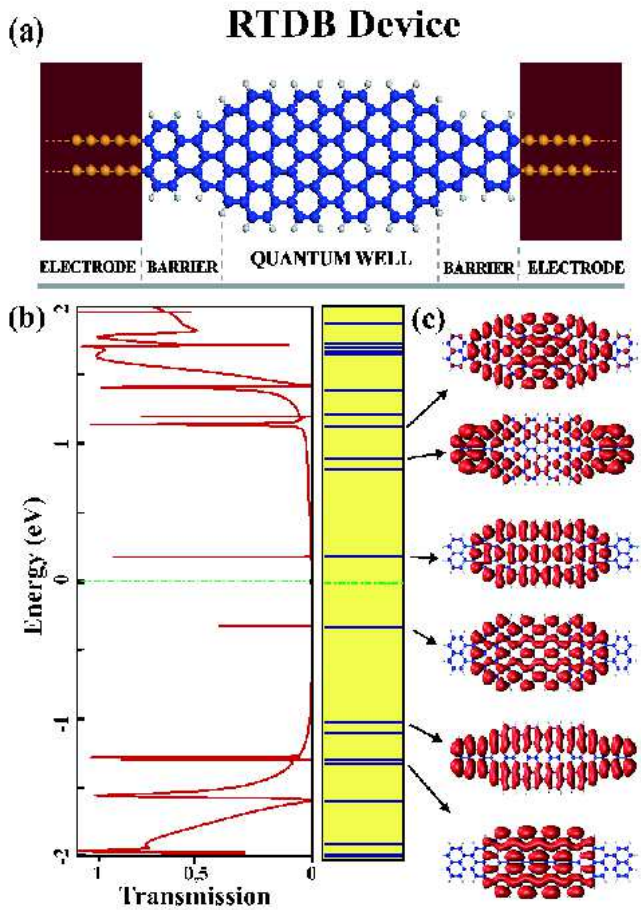}
\caption{(a) Atomic structure of a resonant tunnelling double
barrier (RTDB) device made from a segment of AGSL(10,18;4,4) 
and connected to two metallic electrodes. (b) Transmission 
coefficient, {\textit{\textbf{T}}} versus energy calculated under zero bias. (c) Energy
level diagram of the RTDB and charge densities of confined and extended 
states.} 
\label{fig:trans}
\end{center}
\end{figure}

Finally, we present an interesting device that can operate as a RTDB made preferably from a segment of AGSL(10,18;4,4). 
Such a device is relevant for applications and uses LUMO and HOMO states confined in the wide region. 
For the sake of illustration we considered fictitious metallic electrodes of two widely separated (weakly coupled) carbon chains\cite{tongay} altogether having 4 quantum conductance
channels. 
Six principal layers of electrodes are included at both sides of RTDB as parts of the central device. 
Metallic electrodes make perfect contacts with the central RTDB device. 
Transmission coefficient, {\textit{\textbf{T}}} calculated self-consistently using non-equilibrium Green's Function method\cite{trans} reflects the combined electronic structure of central RTDB device, electrodes and their contacts.
Double-$\zeta$ plus polarization numerical orbitals have been used as the basis set and atomic structures are further optimized before transport calculations.
Fig.~\ref{fig:trans} presents the calculated transmission curve.
Confined LUMO and HOMO states and other confined states 
identified through the energy level diagram and isosurface charge density plots 
give rise to sharp peaks originating from resonant tunnelling effect.
States extending to whole RTDB are coupled with the states of electrode are shifted and contributed broad structures.

In conclusion, various types of quantum structures have been revealed
through size and composition modulation of graphene based nanoribbons. 
The confinement of electron and hole states are shown to lead to interesting effects for new device applications based on graphene.
Several structural and compositional parameters to monitor electronic and transport properties are expected to make these quantum structures an attractive field of research.

We acknowledge fruitful discussions with Dr. E. Durgun. Part of the
computations have been carried out by using UYBHM at
Istanbul Technical University.


\begin{thebibliography}{99}
\bibitem{novo}
Novoselov, K. S.; 
Geim, A. K.; 
Morozov, S. V.; 
Jiang, D.; 
Zhang, Y.; 
Dubonos, S. V.; 
Grigorieva, I. V.; 
Firsov, A. A.; 
Science \textbf{2004}, \textit{306}, 666.

\bibitem{zhang}
Zhang, Y.; 
Tan, Y.-W.; 
Stormer, H. L.; 
Kim, P.; 
Nature \textbf{2005}, \textit{438}, 201.

\bibitem{berger}
Berger, C.; Song, Z.; Li, X.; Wu, X.; Brown, N.; Naud, C.; Mayou, D.; Li, T.; Hass, J.; 
Marchenkov, A. N.; Conrad, E. H.; First, P. N.; de Heer, W. A.; 
Science \textbf{2006}, \textit{312}, 1191.

\bibitem{kats}
Katsnelson, M. I.;
Novoselov, K. S.; 
Geim, A. K.;
Nature Physics \textbf{2006}, \textit{2}, 620.

\bibitem{geim1}
Novoselov, K. S.;
Geim, A. K.;
Morozov, S. V.;
Jiang, D.;
Katsnelson, M. I.;
Grigorieva, I. V.;
Dubonos, S. V.;
Firsov, A. A.;
Nature \textbf{2005}, \textit{438}, 197.

\bibitem{geim2}
Geim, A. K.;
Novoselov, K.S.;
Nature Materials, \textbf{2007}, \textit{6}, 183.

\bibitem{ozyilmaz}
Han, M. Y.;
\"Ozyilmaz, B.;
Zhang, Y.;
Kim, P.;
Phys. Rev. Lett. \textbf{2007}, \textit{98}, 206805.

\bibitem{cohen_hm}
Son, Y.-W.;
Cohen, M. L.;
Louie, S. G.;
Nature \textbf{2006}, \textit{444}, 347.

\bibitem{ribbon1}
Fujita, M.;
Wakabayashi, K.;
Nakada, K.;
Kusakabe, K.;
J. Phys. Soc. Jpn. \textbf{1996}, \textit{65}, 1920.

\bibitem{ribbon2}
Barone, V.;
Hod, O.;
Scuseria, G. E.;
Nano Lett. \textbf{2006}, \textit{6}, 2748.

\bibitem{ribbon3}
Rudberg, E.;
Saek, P.;
Luoi, Y.;
Nano Lett. \textbf{2007}, \textit{7}, 2211.

\bibitem{cohen_gap}
Son, Y.-W.;
Cohen, M. L.;
Louie, S. G.;
Phys. Rev. Lett. \textbf{2006}, \textit{97}, 216803.

\bibitem{dai}
Li, X.;
Zhang, L.;
Lee, S.;
Dai, H.;
Science \textbf{2008}, \textit{319}, 1229.

\bibitem{vasp1}
Kresse, G.;
Hafner, J.;
Phys. Rev. B \textbf{1993}, \textit{47}, 558.

\bibitem{vasp2}
Kresse, G.;
Furthm\"{u}ller, J.;
Phys. Rev. B \textbf{1996}, \textit{54}, 11169.

\bibitem{dft1}
Kohn, W.;
Sham, L. J.;
Phys. Rev. \textbf{1965}, \textit{140}, A1133.

\bibitem{dft2}
Hohenberg, P.;
Kohn, W.;
Phys. Rev. B \textbf{1964}, \textit{76}, 6062.

\bibitem{paw}
Bl\"ochl, P. E.;
Phys. Rev. B  \textbf{1994}, \textit{50}, 17953.

\bibitem{pw91}
Perdew, J. P.;
Chevary, J. A.;
Vosko, S. H.;
Jackson, K. A.;
Pederson, M. R.;
Singh, D. J.;
Fiolhais, C.;
Phys. Rev. B \textbf{1992}, \textit{46}, 6671.


\bibitem{cohen_gw}
Yang, L.;
Park, C.-H.;
Son, Y.-W.;
Cohen, M. L.;
Louie, S. G.;
Phys. Rev. Lett. \textbf{2007}, \textit{99}, 186801.

\bibitem{tekman}
Tekman, E.;
Ciraci, S.;
Phys. Rev. B \textbf{1989}, \textit{40}, 8559.

\bibitem{esaki}
Esaki, L.;
Chang, L. L.;
Phys. Rev. Lett. \textbf{1974}, \textit{33}, 495.


\bibitem{bastard}
Bastard, G.;
\textit{``Wave mechanics applied to semiconductor heterostructures''},
Les Editions de Physique, Les Ullis, France, \textbf{1988}.

\bibitem{ciraci1}
Ciraci, S.;
I. P. Batra, 
Phys. Rev. Lett. \textbf{1987}, \textit{58}, 1982.

\bibitem{ciraci2}
Ciraci, S.;
Batra, I. P.;
Phys. Rev. B \textbf{1987}, \textit{36}, 1225.

\bibitem{ciraci3}
Ciraci, S.;
Batra, I. P.;
Phys. Rev. B \textbf{1988}, \textit{38}, 1835.

\bibitem{ciraci4}
Ciraci, S.;
Baratoff, A.;
Batra, I. P.;
Phys. Rev. B \textbf{1990}, \textit{41}, 2763.



\bibitem{tongay}
Tongay, S.;
Senger, R. T.;
Dag, S.;
Ciraci, S.;
Phys. Rev. Lett. \textbf{2004}, \textit{93}, 136404.

\bibitem{trans}
Brandbyge, M.;
Mozos, J.-L.;
Ordejón, P.;
Taylor, J.;
Stokbro, K.;
Phys. Rev. B \textbf{2002}, \textit{65}, 165401.

\end{thebibliography}
\end{document}